\documentclass{article}

\usepackage[nonatbib,final]{neurips_2022}

\usepackage[utf8]{inputenc} 
\usepackage[T1]{fontenc}    
\usepackage{hyperref}       
\usepackage{url}            
\usepackage{booktabs}       
\usepackage{amsfonts}       
\usepackage{nicefrac}       
\usepackage{microtype}      
\usepackage{xcolor}         

\usepackage[english]{babel}
\usepackage{csquotes}
\usepackage{physics}
\usepackage{mathtools}
\usepackage{isomath}
\usepackage{float}
\usepackage{wrapfig}
\usepackage{multirow}
\usepackage{etoolbox}
\usepackage[backend=biber, style=phys, sorting=none]{biblatex}
\usepackage{subcaption}
\usepackage{chngcntr}

\usepackage{amsthm}

\title{A physics-informed search for metric solutions to Ricci flow, their embeddings, and visualisation}

\author{%
  Aarjav Jain \\
  Dept. of Physics\\
  Cambridge University\\
  \texttt{aj562@cam.ac.uk} \\
  \And
   Challenger Mishra \\
   Dept. of Computer Science\\
   Cambridge University\\
  \texttt{cm2099@@cam.ac.uk} \\
  \And 
  Pietro Li{\`o} \\
   Dept. of Computer Science\\
   Cambridge University\\
  \texttt{pl219@cam.ac.uk} \\
}

\begin{document}

\maketitle

\begin{abstract}
  Neural networks with PDEs embedded in their loss functions (physics-informed neural networks) are employed as a function approximators to find solutions to the Ricci flow (a curvature based evolution) of Riemannian metrics. A general method is developed and applied to the real torus. The validity of the solution is verified by comparing the time evolution of scalar curvature with that found using a standard PDE solver, which decreases to a constant value of 0 on the whole manifold. We also consider certain solitonic solutions to the Ricci flow equation in two real dimensions. 
We create visualisations of the flow by utilising an embedding into $\mathbb{R}^3$. Snapshots of highly accurate numerical evolution of the toroidal metric over time are reported. We provide guidelines on applications of this methodology to the problem of determining Ricci flat Calabi--Yau metrics in the context of String theory, a long standing problem in complex geometry.
\end{abstract}

\section{Introduction}
Ricci flow is a powerful tool for geometric analysis \cite{chow2006hamilton}. Formulated by Hamilton in the 80's, Grigori Perelman employed Ricci flow to resolve the  Poincar\'e conjecture in 2002, proving that every simply connected, closed manifold in three dimensions is \textit{homeomorphic} to the 3-sphere \cite{perelman2002entropy}. Intuitively, Ricci flow is a geometric flow of a metric on a smooth Riemannian manifold, which uniformizes curvature of the metric. Given a smooth Riemannian manifold {$\mathcal{M}$}, Ricci flow assigns a metric \emph{g} to each \emph{time} $t$, dictated by the Ricci tensor of the metric, \textit{Ric(g)}~\cite{sheridan}. Mathematically put, 
\begin{equation}\label{RF}
\partial_{t} g_{a b}(x,t) = -2Ric_{a b}(g(x,t)),~\text{where~$a$, $b$}\in\{1,2,\dots,dim({\mathcal{M}})\}.
\end{equation}
Ricci flow has deep connections to \textit{renormalization group flow}, the latter appearing in the context of \textit{beta functions} in quantum field theories, which formulate the dependency of coupling constants on energy scales \cite{bakas2007renormalization, carfora2010renormalization}. The flow parameter $t$ in Equation~\ref{RF} is reminiscent of such scales. Following decades of further advances, Ricci flow has found connections with training in neural networks, as well as in community learning~\cite{glass2020riccinets} and studying robustness under gradient descent perturbations~\cite{chen2021thoughts}.

Our interest in Ricci flow concerns complex geometries which appear as extra dimensions in certain superstring theories. A class of geometries of particular interest is \textit{Calabi--Yau} manifolds. One problem of particular interest is related to the Calabi conjecture, which postulated the existence of metrics over such geometries that have a vanishing Ricci tensor, that is, `Ricci flat'. This amounts to finding the fixed point of the PDE in Equation~\ref{RF}- an analytic form of which remains elusive till date. Further work in the interim resulted in numerical treatments leading to approximation techniques \cite{headrick2005numerical} including the celebrated Donaldson's algorithm \cite{donaldson}. Recently, machine learning was demonstrated to be a powerful tool in the context of approximating Ricci flat Calabi--Yau metrics for complex geometries with special holonomy~\cite{mishra2022neural, douglas2022numerical, anderson2021moduli}. These approaches rely on neural network approximations of the metric of interest. This is reminiscent of physics inspired neural networks (PINNS)\cite{raissi2017physics}; albeit the only data in the machine learning pipeline involves points sampled from such geometries. Hinted by \cite{mishra2022neural}, in this work we propose to formulate Ricci flow on real geometries using PINNs. We focus on the real torus, and consider cigar solitonic solutions to the Ricci flow equation~\ref{RF}. 
\section{Using neural networks to learn Ricci flow}
\label{NN_RF}
The flow described in \ref{RF}, is the  `unnormalised' Ricci flow. As such the volume of the given Riemannian manifold can evolve with time. The action of the flow is to uniformize curvature over the surface, in a way analogous to the heat equation spreading heat evenly over space. The action of Ricci flow on the round metric of a sphere can be shown to shrink the sphere to a point \cite{sheridan}. We utilise the Nash--Kuiper theorem to embed the geometries in~$\mathbb{R}^{3}$. We propose to use Physics-informed neural networks (PINNs) to solve the Ricci flow PDE for two-dimensional real manifolds. PINNs incorporate the PDEs, and the key symmetry/initial conditions of the system into the loss function, meaning that once trained, the neural network will output a solution to the PDE (\cite{PINN1}). The metric $g{(x,t)}$, on the manifold $\mathcal{M}$, is approximated by the neural network $g_\theta(x,t)$, where $\theta$ denotes the parameters of the neural network. The PDE residual is calculated from the Ricci flow equation \ref{residual} using the neural network approximation. From this, we define a residual loss function, $\mathcal{L}_{res}{(\theta)}$ in Equation \ref{residual}. Similarly, residual losses associated with the required initial and symmetry conditions are constructed. These are $\mathcal{L}_{init}{(\theta)}$ and $\mathcal{L}_{sym}{(\theta)}$ respectively, and they are combined to make the PINN loss function in Equation \ref{loss_func}. The $\mathcal{L}_{sym}{(\theta)}$ loss can include terms which mandate a desired symmetry. If we know the initial metric has rotational symmetry, we could enforce this by inclusing a rotational symmetry loss function. More on this in Section \ref{experiments}.
\begin{eqnarray}\label{residual}
\mathcal{L}_{res}{(\theta)}:=\frac{1}{N} \sum ^{N}_{i=1}\left| r_{\theta}(t_i,x_i )\right|^2, \text{~with~~}
r_\theta(t,x):= \frac{\partial}{\partial t}g_\theta(t, x) + 2Ric(g_\theta(t,x))\text{~,~} x_i \in \mathcal{M}. 
\end{eqnarray}
\begin{equation}\label{loss_func}
\mathcal{L}{(\theta)} = \mathcal{L}_{res}{(\theta)}+\lambda_{0}\ \mathcal{L}_{\text{init}}{(\theta)}+\lambda_1\  \mathcal{L}_{\text{sym}}{(\theta)}.
\end{equation}
\section{Experiments}
\label{experiments}
We apply the PINN approach to Ricci flow of a metric on the real 2-dimensional torus. We also consider cigar solitonic solutions of Ricci flow as an additional study, since this has analytic solutions to Ricci flow, hence allows the benchmarking  of PINN generated solutions. 

\begin{figure}%
    \centering
    \subfloat{{\includegraphics[width=0.4\linewidth ,height=0.25\linewidth]{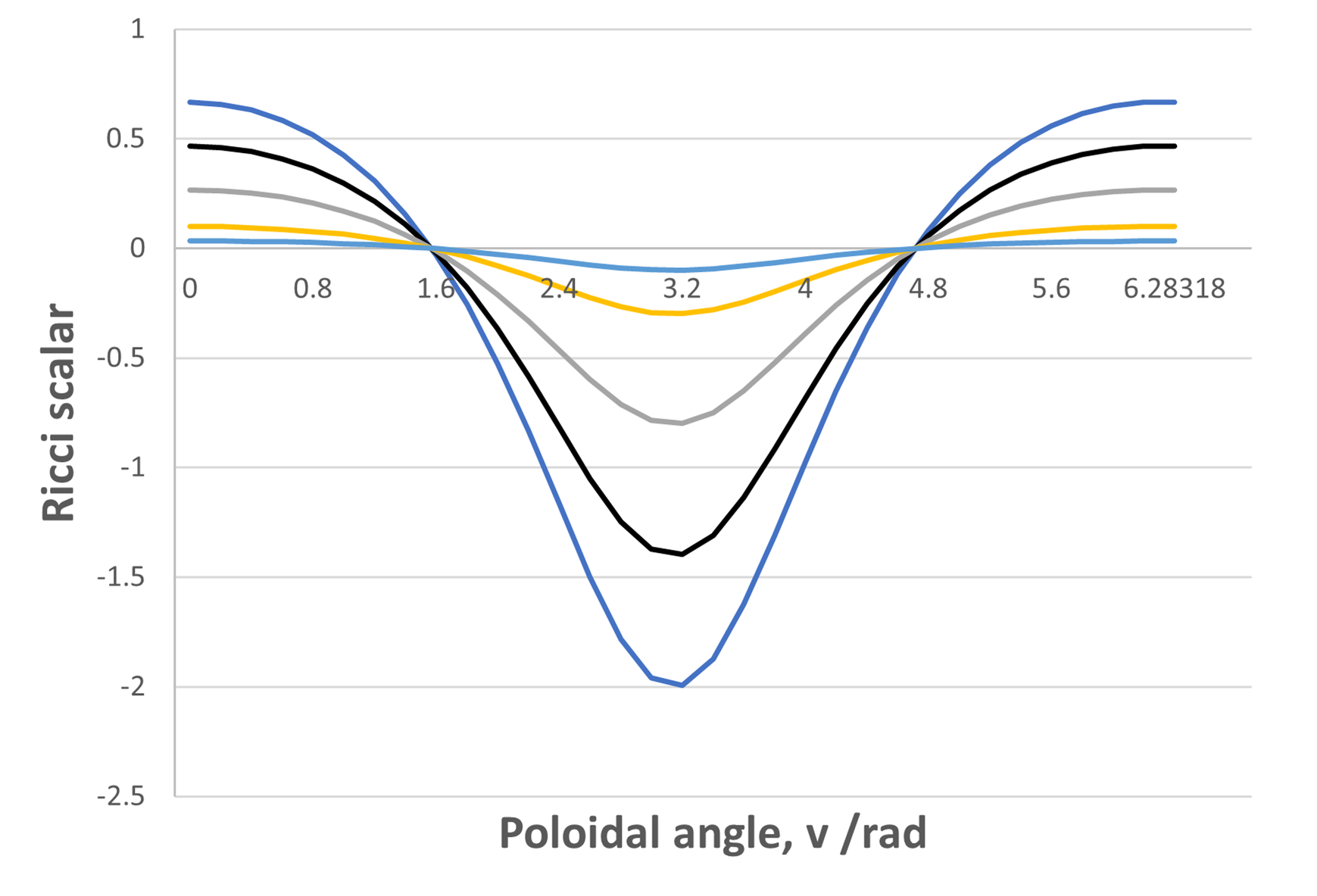} }}%
    \qquad
    \subfloat{{\includegraphics[width=0.4\linewidth ,height=0.25\linewidth]{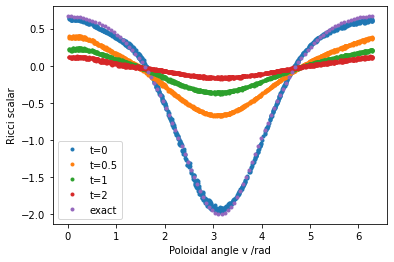} }}%
    \caption{Evolution of the scalar curvature under Ricci flow. 
    A Mathematica PDE solver generated the left figure; the right corresponds to our PINN solution. It captures the initial conditions, and shows accurate long time behaviour of vanishing scalar curvature, the fixed point of Ricci flow.}%
    \label{fig:T2_RF}%
\end{figure}

\begin{figure}[t]
\begin{subfigure}{.31\textwidth}
  \centering
  \includegraphics[width=.85\linewidth]{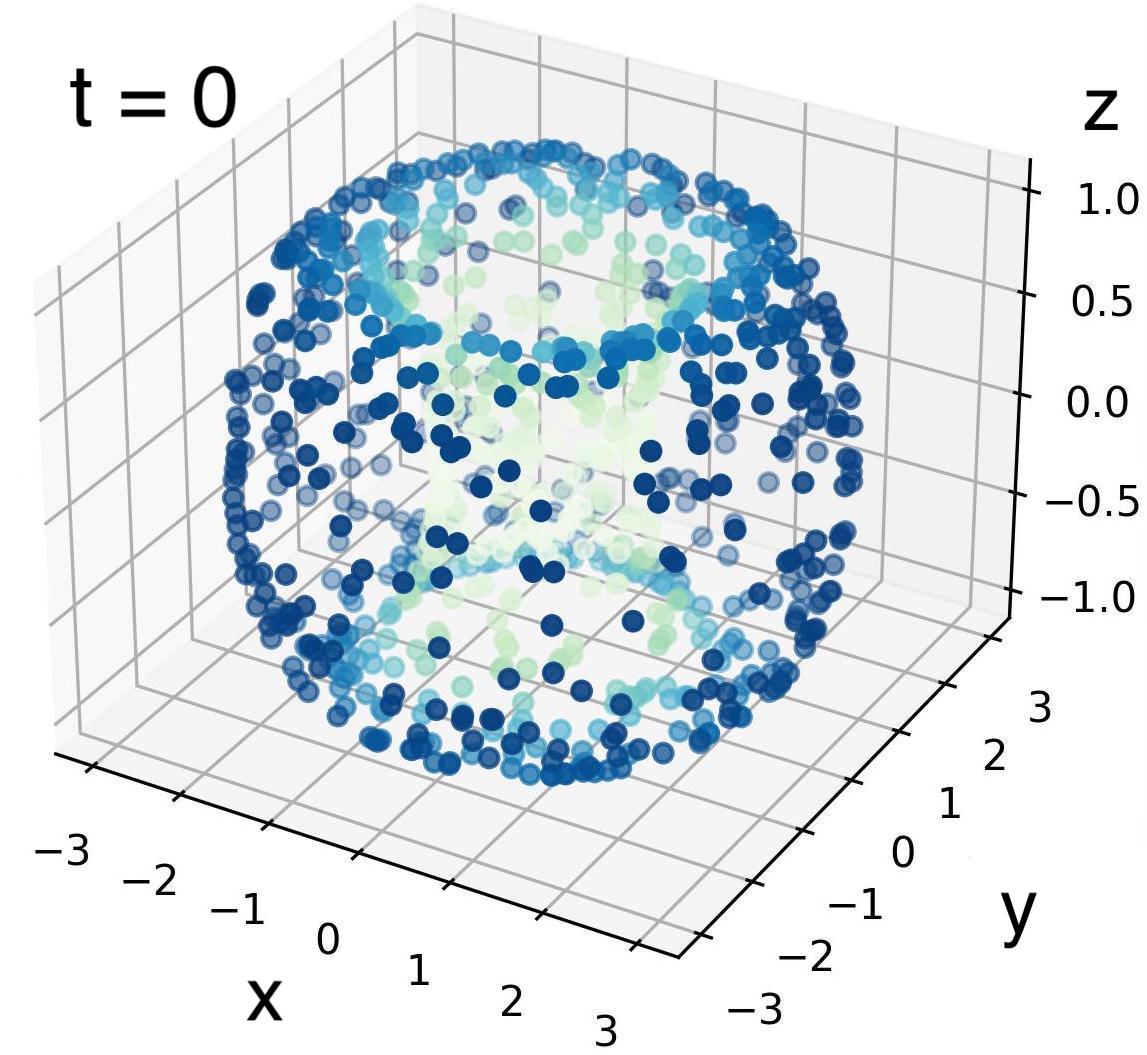}
  \label{fig:sfig1}
\end{subfigure}%
\begin{subfigure}{.31\textwidth}
  \centering
  \includegraphics[width=.85\linewidth]{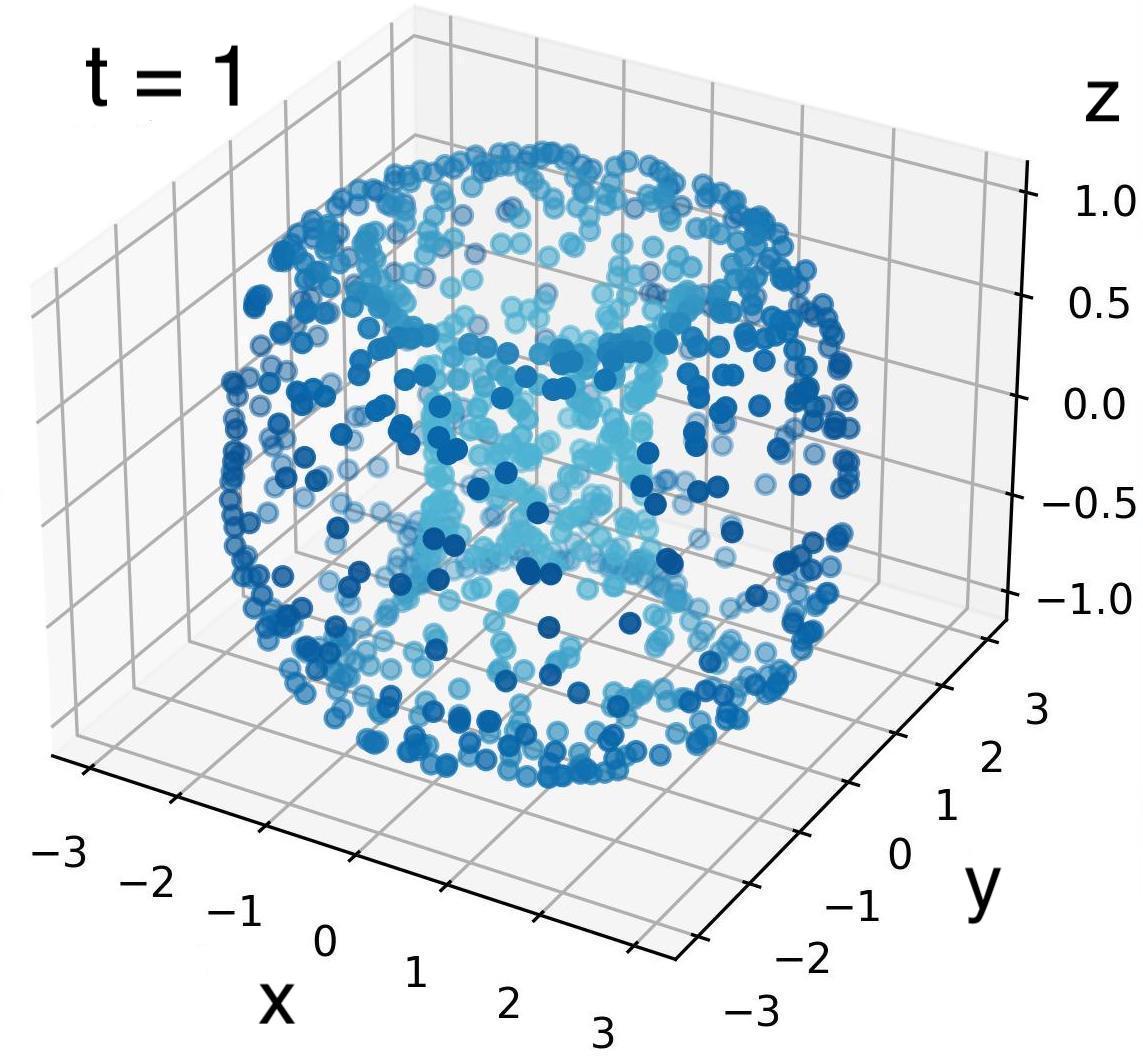}
\end{subfigure}
\begin{subfigure}{.38\textwidth}
  \centering
  \includegraphics[width=.9\linewidth]{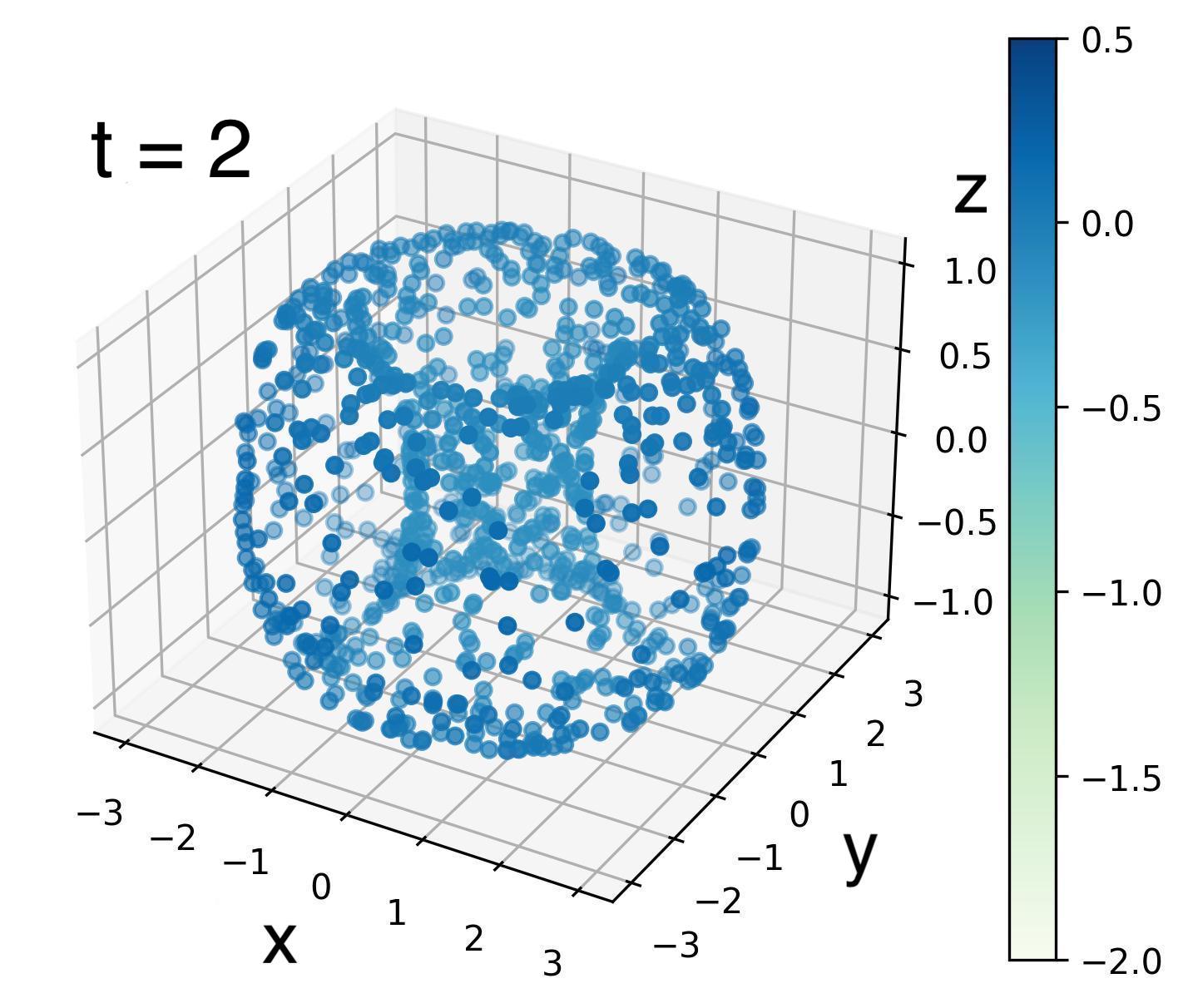}
\end{subfigure}
\caption{Colour plot of Ricci scalar of a torus evolving under Ricci flow. Here, unlike the embedding case, the base geometry is kept the same, showcasing the decay of curvature with Ricci flow.}\label{fig:colour_plot}
\end{figure}

\textbf{Real Torus:~~} The torus is spanned by toroidal and poloidal angles, denoted by $u$ and $v$ respectively, with $u\in[0,2\pi)$ and $v\in[0,2\pi)$.
An illustration of this parameterisation is in the Figure \ref{torus_coord} in the Appendix~\ref{appendix}. 
A natural metric tensor on the real torus is,
\begin{eqnarray}\label{torus metric}
g(t, u, v)=
\begin{pmatrix}
(c+a\cos v)^{2} & 0 \\ 0 & a^{2}
\end{pmatrix},\text{~~with symmetry conditions, }\\
g(t,u,v) = g(t,u+\delta,v) = g(t,u,2\pi-v), \text{  where   } \delta\in[0,2\pi)\label{periodic_BC},
\end{eqnarray}
which follow from the symmetries of the torus. Training is done using the Adam optimiser \cite{adam} on a fully connected feed-forward network with three hidden layers of 16, 32 and 16 units, and Softplus activation. 
We use 1000 randomly generated training points in the domain. Our implementation is in JAX which supports efficient gradient computations, through auto differentiation. This is particularly important in light of the fact that the Ricci tensor, $Ric(g)$, is dependent on the derivatives of the metric function. In Figures \ref{fig:T2_RF} and \ref{fig:colour_plot}, we show the evolution of the scalar curvature (which is the trace of the $Ric(g)$) with the flow parameter. We compare the PINN generated solution with numerical simulations using a standard PDE solver in Mathematica. We choose $c$ = 2 and $a$ = 1. The solutions bear significant resemblance to each other. The initial Ricci scalar curve matches the exact result which can be found by taking derivatives of the initial torus metric shown in Equation \ref{torus metric}- a useful sanity check. The PINN and exact results have an MSE of $\sim 1.8\times10^{-3}$. The computation took around 4 hours on an i5 laptop. An additional visualisation is provided in Figure \ref{fig:colour_plot}, where the torus base geometry is kept the same, but the decay of curvature with time under Ricci flow is depicted with colour representing scalar curvature. Curvature decays to a zero with time, as expected.
\begin{figure}[t]
    \centering
  {{\includegraphics[width=0.4\linewidth,height=0.35\linewidth]{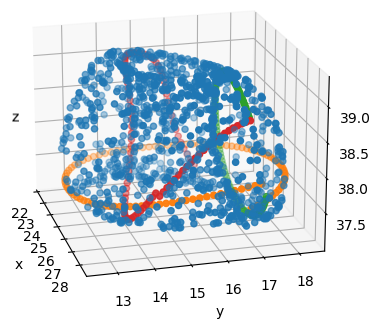} }}%
    \qquad
    {{\includegraphics[width=0.4\linewidth,height=0.32\linewidth]{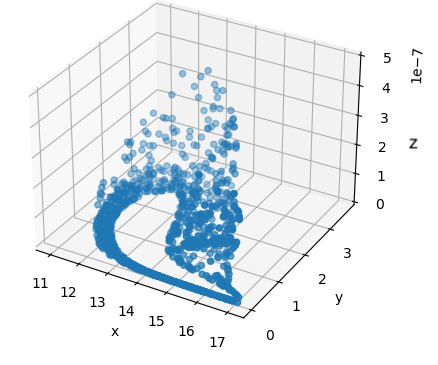} }}%
    \caption{Left is a plot of the initial torus metric. Its geodesics are plotted in colours. Right is a plot of the torus  Ricci flow ($t=2$). Note the scale in the $z$ direction. It does not have the required rotational symmetry, but has been severely flattened in the $z$ direction, as expected.}%
    \label{fig:metric_plotting}%
\end{figure}
\begin{figure}
    \centering
    \subfloat{{\includegraphics[width=0.4\linewidth,height=0.35\linewidth]{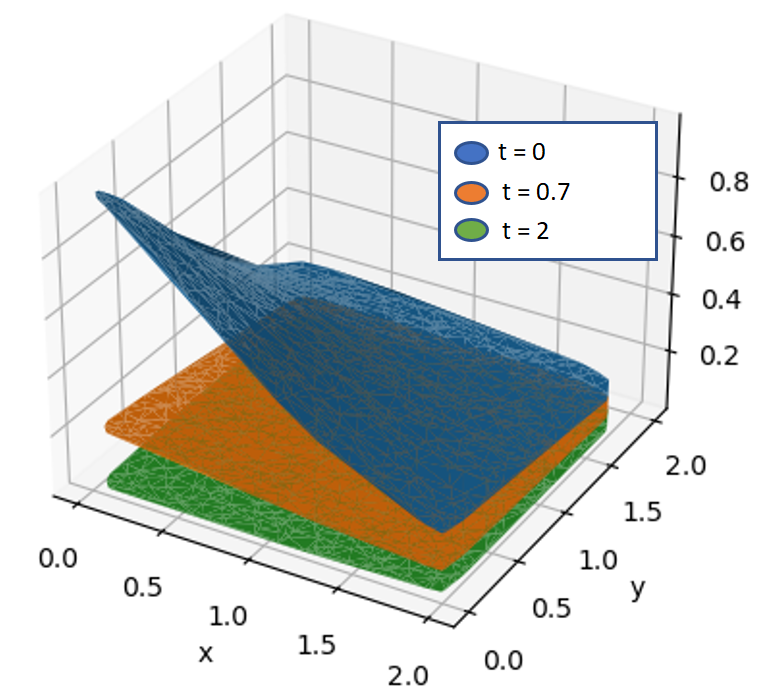} }}%
    \qquad
    \subfloat{{\includegraphics[width=0.4\linewidth,height=0.3\linewidth]{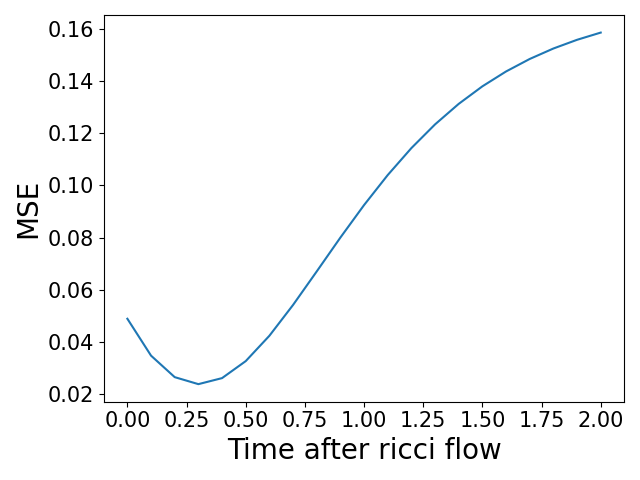} }}%
    \caption{The figure on the left is the $g_{11}=g_{22}$ component of the Cigar soliton solution generated by PINN under Ricci flow. This compares favourably with the analytic solution in Equation \ref{cigarsolitonsolution}. On the right we show the MSE. It is $\mathcal{O}(10^{-2})$ at small times, and grows under Ricci flow.}%
    \vspace{-15pt}
    \label{cigar decay}%
\end{figure}
\\\textbf{Cigar Soliton:~~}
An example of a two dimensional metric with an analytic solution to the Ricci flow is the cigar soliton \cite{topping}. It has metric $g(\textbf{x})=({1+x^2+y^{2}})^{-1}\mathbb{I}_2$, with its Ricci flow solution given by:
\begin{equation}\label{cigarsolitonsolution}
    g(t,\textbf{x})=({e^{4t}+x^2+y^{2}})^{-1}\mathbb{I}_2, \text{~where,~~~} (x,y)\in \mathbb{R}^{2}.
\end{equation}
The PINN solution approximates the analytic metric for all values of the flow parameter, $t$.
A three dimensional plot of the $g_{11}=g_{22}$ component as a function of $x$ and $y$ is shown on the left of Figure \ref{cigar decay}.
The behaviour of the PINN generated solution bears close resemblance to the analytic solution, although over time, the MSE gradually grows. 
We expect high accuracies at lower times since the residuals from the metric at $t=0$ is minimised by choice of loss function (Equation \ref{loss_func}). The minima occurs since, as time increases, the metric values themselves decrease, and this is overcome at higher time.
\section{Metric plotting}
\label{viz}
We aim to plot the surface corresponding to the learned neural network solution to Ricci flow, $g_{\theta}(t,u,v)$. That is, we wish to create an embedding of the manifold into $\mathbb{R}^3$, and plot this for a number of randomly generated points to give a three dimensional plot of the manifold for multiple time durations under Ricci flow. 
A smooth embedding of the geometries we have considered into Euclidean space exists ala the Nash-Kuiper embedding theorem. We denote this by $f:M\rightarrow \mathbb{R}^{n}$. The embedding is given by a system of first-order PDEs:
\begin{equation}
\label{Nash}
 g_{ij}(x)=\sum _{\alpha =1}^{n}{\frac {\partial f^{\alpha }}{\partial x^{i}}}{\frac {\partial f^{\alpha }}{\partial x^{j}}},     \alpha \in 1,2,\dots, n
\end{equation}
where index $\alpha$ runs over the number of dimensions for the target embedding (3 in this case). Embedding into $\mathbb{R}^{3}$ leaves a system of the three PDEs (An explicit unpacking is detailed in Appendix~\ref{appendix}), the solution of which give $f^{i}(u,v)$, where $i = dim(\mathcal{M})$. This embedding PDE is solved using PINNs. The architecture, epochs and training splits are identical to Section \ref{experiments}. The training data consists of 1000 data points with random angles on the torus at a fixed time point, with the output being the corresponding points in $\mathbb{R}^{3}$ (embedding).
We proceed by sampling random points on the base metric $(u_i,v_i)$ and applying the embedding functions $f^{i}$ to each of these, as
\begin{equation}\label{embedding_coords}
(x_i,y_i,z_i) = (f^{x}(u_i,v_i),f^{y}(u_i,v_i),f^{z}(u_i,v_i)).
\end{equation}
These points in $\mathbb{R}^{3}$ can be plotted to give a visualisation of the manifold at different times during Ricci flow. Ricci flow is diffeomorphism invariant, so the symmetries of the initial metric should be preserved. Hence, symmetries in Equation \ref{periodic_BC} are included in the loss function through $\mathcal{L}_{\text{sym}}$ in Equation~\ref{residual}. By embedding 1000 random points on the torus into $\mathbb{R}^{3}$ and plotting them we get plots for initial torus, and torus after 2 seconds of Ricci flow, seen in Figure \ref{fig:metric_plotting}. The geodesic equation is solved for trivial initial directions using this embedding, and the curves are plotted on the surface of the initial torus, giving the expected equatorial paths. Training took 8 hours on an i5 laptop. After two seconds, the manifold is much flatter, which is expected (notice the scale in $z$-direction). In the figure on the right, the expected rotational symmetry is not seen. It was found that this plot depended heavily on the regularisation terms applied to various loss terms (see Equation \ref{residual}). In this experiment, fixed scalars were used for these terms, but if an adaptive scheme is used (such as \cite{adaptive_loss}), then a term corresponding to rotational symmetry could be enhanced when necessary during training.  
\section{Discussion and Limitations}
\label{discussion}
We applied a PINN technique to solve the Ricci Flow PDE for real, two-dimensional geometries. The solution for a torus matched well to that found by a PDE solver, and had an MSE of $\mathcal{O}(10^{-3})$ compared with the exact solution for the toroidal initial metric. The cigar soliton geometry matched the analytic solution, with an MSE of $\mathcal{O}(10^{-2})$. The learned PINN metric for torus Ricci flow was used to solve PDEs derived from the Nash--Kuiper Embedding theorem using PINNs. This allows visualisations of manifolds under Ricci Flow, where the geodesics of the metric are considered as spanning the base geometry. These showed elements of correctness; a toroidal figure was seen before Ricci Flow, and a flatter manifold after Ricci Flow. Full rotational symmetry was not perfectly replicated in either case, which are potentially due to compounding numerical errors during the course of Ricci flow, although the growth shows signs of stabilising. Instead of demanding rotational symmetry from the loss function, one may embed it directly with an appropriate coordinate system. This was avoided here, since choosing one can become difficult when working with larger dimensional and/or complex manifolds.
Although deliberate in our choice of architecture, the approach would benefit from the use of equivariant architectures. Further, the magnitudes of various loss terms being important to the final embedding, an adaptive technique such as \cite{adaptive_loss} could be employed.
The PINN approach allowed simple differentiation of the neural network solution, to find relevant quantities of Ricci tensor and curvature, as well as being beneficial in the context of Nash--Kuiper embedding. 
The results presented show the promise of using PINN techniques in Ricci flow of higher dimensional manifolds and complex manifolds, and could aid in their visualisation. The tool we have created is completely general, and allows adaptation to the setting of real manifolds by simply changing the initial metric loss function. 

Embeddings of complex analogue of the real torus, K3 surfaces, and Calabi--Yau manifolds
are possible, when technical aspects related to sampling and constructing meaningful loss functions~\cite{douglas2022numerical, anderson2021moduli,mishra2022neural} are considered. See~\ref{appendix} for further details. Solutions to Ricci flow have substantial applications in superstring theory wherein extra dimensional geometries play a crucial role in determining four-dimensional physics, enabling predictions of particle masses and coupling constants from string theory for the first time, which rely on an accurate computation of the Ricci flat metric. 
\section{Broader Impact}
Ricci flow is mathematically rich and a powerful tool for geometric analysis
We develop and demonstrate a framework for learning solutions to Riemannian metrics, and provide special cases for real 2$D$ geometries. This framework can be extended to complex geometries with increasing dimension, such as the complex torus, K3 surfaces, and Calabi--Yau manifolds, which have applications in superstring theory: enabling prediction of particle masses. We also presenet an embedding framework of geometries through PINNS and this idea can aid visualisation of higher dimensional manifolds.


\printbibliography

\appendix
\section{Appendix}\label{appendix}
In this appendix we provide some further details on the parameterisation of the real torus, and an explicit form of the partial differential equations that appear in the Nash-Kupier Embedding theorem. 
\subsection{Embeddings and Ricci flow}

\begin{figure}[h]
	\centering
	\includegraphics[height=0.3\textwidth]{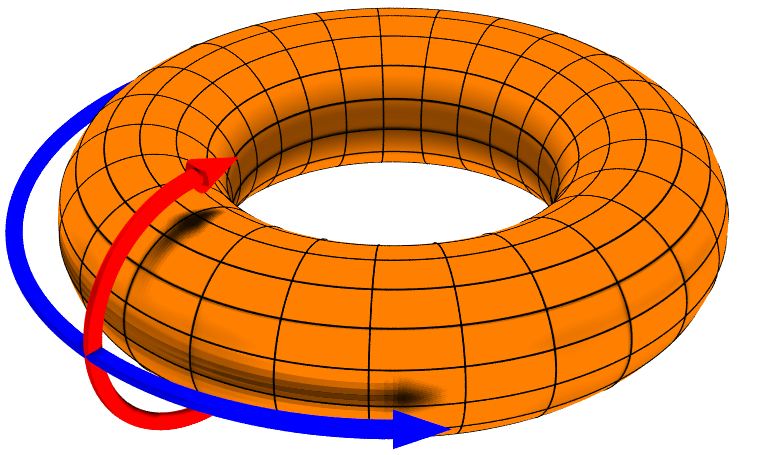}
	\caption{A diagram depicting the the toroidal ($u$) direction, represented by the blue arrow, and the poloidal ($v$) direction, represented by the red arrow}
	\label{torus_coord}
\end{figure}

\subsubsection*{Torus curvature decay}
During Ricci flow, the scalar curvature $R$ of the torus is expected to reduce with time. Figure \ref{fig:colour_plot} shows stills of an animation produced to showcase this. The animation can be found in the GitHub link.  

\subsubsection*{Nash-Kupier Embedding equation} 
\begin{equation}
\label{Nash}
 g_{ij}(x)=\sum _{\alpha =1}^{n}{\frac {\partial f^{\alpha }}{\partial x^{i}}}{\frac {\partial f^{\alpha }}{\partial x^{j}}},     \alpha \in 1,2,\dots, n
\end{equation}

\subsection{Complete Intersection Calabi--Yau manifolds}
Calabi--Yau manifolds are complex, \textit{K\"ahler} manifolds with vanishing first \textit{Chern class}, or equivalently, with \textit{holonomy} group $SU(n)$, where $n$ is the complex dimension of the manifold. A large class of Calabi--Yau manifolds, is the Complete Intersection Calabi--Yau manifolds (CICYs)~\cite{candelas1988complete}. Compiled in the late 80's, this is  the first such dataset in algebraic-geometry. The ambient space for these geometries is a product of complex projective spaces (as opposed to ~$\mathbb{R}^{n}$). The embedding is given by the zero locus of intersecting polynomials with degrees that satisfy certain constraints that ensure the Calabi--Yau property~\cite{hubsch1992calabi}. The class of Complete Intersection Calabi--Yau manifolds (CICYs) in three dimensions are a family of 7890 geometries. A CICY $\mathcal{M}$ is described by the vanishing locus of a number of polynomials with co-ordinates in an ambient space $\cal{A}$ which is a product of complex projective spaces of different dimensions. That is, 
\begin{equation*}
 \mathcal{M} \subset {\cal A}=\mathbb{P}^{n_1}\times\cdots\times\mathbb{P}^{n_m}\; .
\end{equation*}
The deformation class of $\mathcal{M}$ is given by a configuration matrix which contains the multi-degrees the defining polynomials. These geometries have an equivalent bipartite graph representation in which the polynomials and components of the projective spaces are represented as distinct types of nodes. The multi-degrees are captured by the number of connections between a given space and a polynomial. A couple of examples follow. 
\begin{center}
\includegraphics[width=4cm]{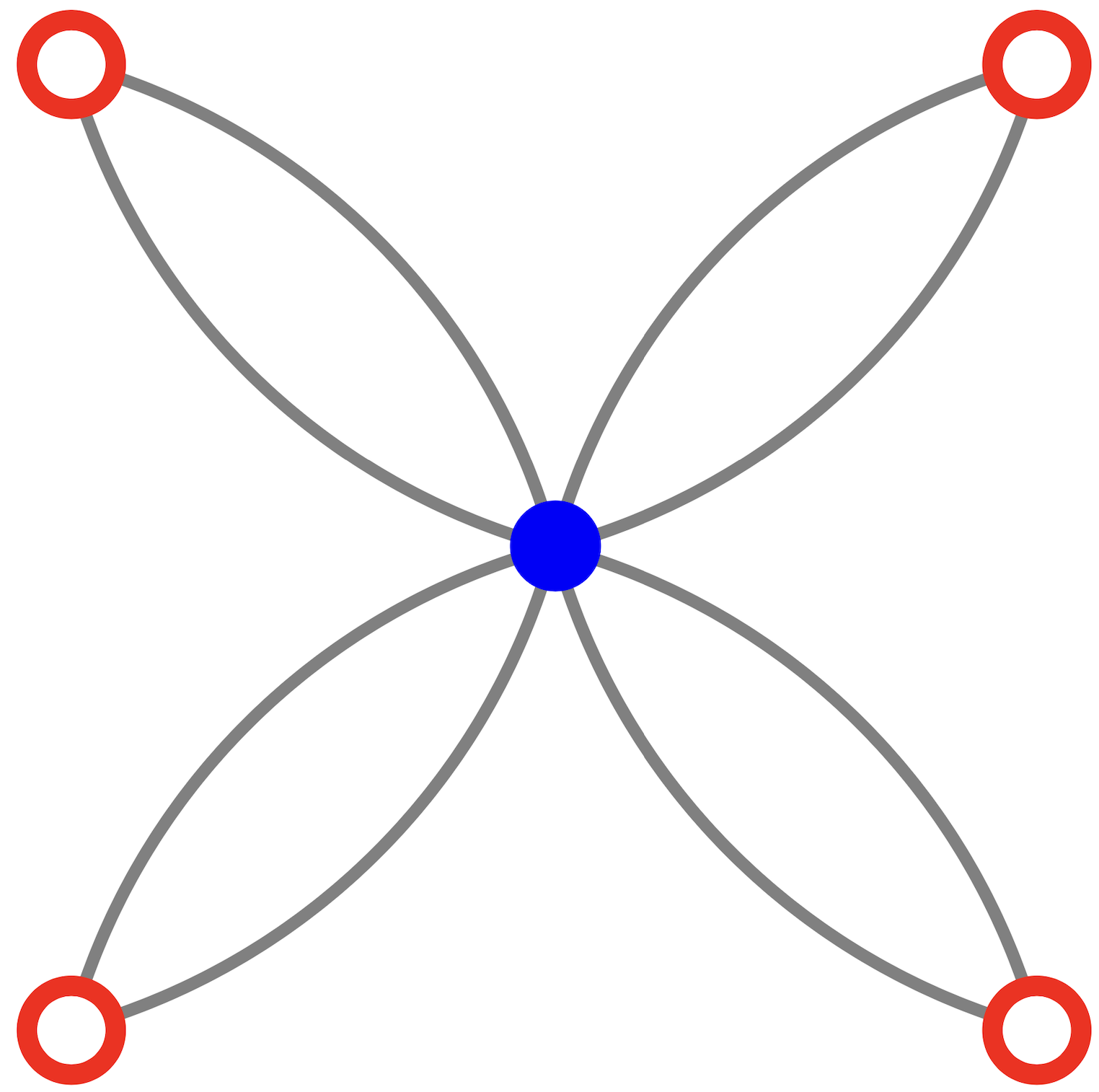}\quad\quad\quad\quad\quad\quad \includegraphics[width=4cm]{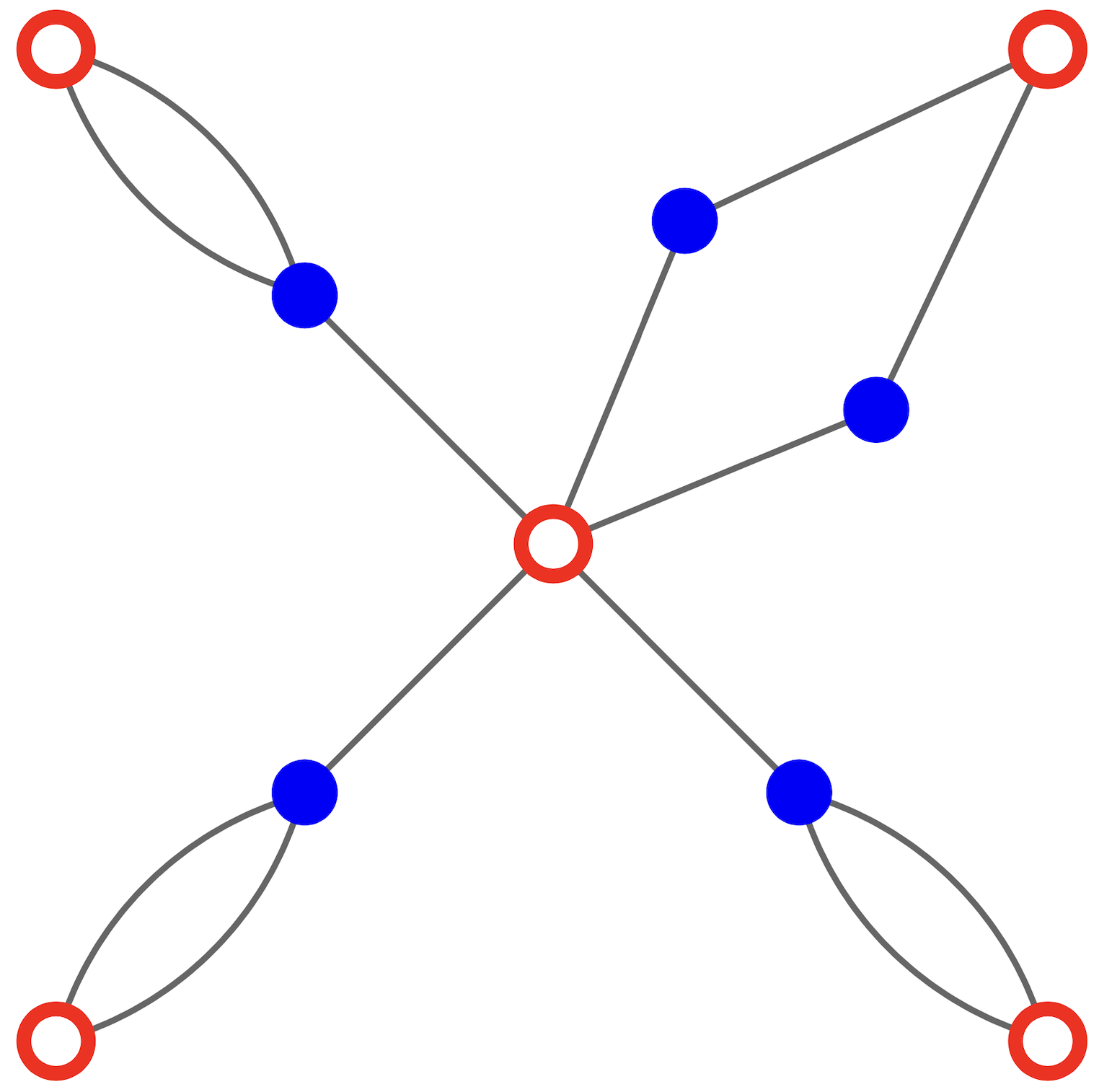}
\end{center}

The bipartite graph on the left, refers to a Calabi--Yau geometry, known as \textit{tetraquadric}. The one to the right is a \textit{split} of this~\cite{hubsch1992calabi}, which is a geometric transition between the two manifolds. It is apparent from these graphical representations that these geometries can admit discrete symmetries~\cite{Candelas:1987du,Candelas:2008wb,Candelas:2010ve,Braun:2010vc,lukas2020discrete, mishra2017calabi,candelas2018highly,mishra2017calabi} and as such are amenable to a treatment developed in this paper using equivariant architectures.



\end{document}